\documentclass[fp,twocolumn]{jpsj3}

\usepackage{txfonts}
\usepackage{graphicx}
\usepackage{amssymb, amsfonts, amsmath}
\usepackage{here}
\usepackage{dcolumn}
\usepackage{bm}
\usepackage{color}
\usepackage{here}
\usepackage{epstopdf}
\usepackage{textcomp}

\makeatletter

\newlength{\figwidth}
\newcommand{\BaRb}{Ba$_{1-x}$Rb$_{x}$Fe$_{2}$As$_{2}$}
\newcommand{\BaK}{Ba$_{1-x}$K$_{x}$Fe$_{2}$As$_{2}$}
\newcommand{\KAs}{KFe$_{2}$As$_{2}$}
\newcommand{\RbAs}{RbFe$_{2}$As$_{2}$}

\newcommand{\BaAs}{BaFe$_{2}$As$_{2}$}
\newcommand{\BaCo}{Ba(Fe$_{1-x}$Co$_x$)$_{2}$As$_{2}$}
\newcommand{\tc}{$T_{{\rm c}}$}

\makeatother

\title{Charge Transport in Ba$_{1-x}$Rb$_{x}$Fe$_{2}$As$_{2}$ Single Crystals}

\author{Masaya Tsujii$^1$,
Kousuke Ishida$^1$\thanks{kousuke.ishida@cpfs.mpg.de}\thanks{Present address: Max Planck Institute for Chemical Physics of Solids, N\"{o}thnitzer Stra{\ss}e 40, 01187 Dresden, Germany.},
Shigeyuki Ishida$^2$,
Yuta Mizukami$^1$,
Akira Iyo$^2$,
Hiroshi Eisaki$^2$,
Takasada Shibauchi$^1$\thanks{shibauchi@k.u-tokyo.ac.jp}
}

\inst{$^1$Department of Advanced Materials Science, University of Tokyo, Kashiwa, Chiba 277-8561, Japan \\
$^2$ National Institute of Advanced Industrial Science and Technology (AIST), Tsukuba, Ibaraki 305-8568, Japan} 

\date{\today}

\abst
{Recent studies in heavily hole-doped iron-based superconductor RbFe$_2$As$_2$ have suggested the emergence of novel electronic nematicity directed along the Fe-As direction, 45$^\circ$ rotated from the usual nematicity ubiquitously found in BaFe$_2$As$_2$ and related materials. 
This motivates us to study the physical properties of \BaRb, details of which remain largely unexplored. 
Here we report on the normal-state charge transport in \BaRb \ superconductors by using high-quality single crystals in the range of Rb concentration $0.14\le x \le 1.00$. 
From the systematic measurements of the temperature dependence of electrical resistivity $\rho(T)$, we find a signature of a deviation from the Fermi liquid behavior around the optimal composition, which does not seem related to the antiferromagnetic quantum criticality but has a potential link to hidden nematic quantum criticality. 
In addition, electron correlations derived from the coefficient of $T^2$ resistivity show a marked increase with Rb content near the heavily hole-doped end, consistent with the putative Mott physics near the $3d^5$ electron configuration in iron-based superconductors.}

\begin{document}
\maketitle

\section{INTRODUCTION}
\BaAs-based superconductors are the prototypical family of iron-based superconductors. 
Starting from the antiferromagnetic parent compound \BaAs, superconductivity can be induced by electron or hole doping \cite{sefat2008superconductivity,rotter2008superconductivity}, isovalent substitution \cite{jiang2009superconductivity}, and by applying pressure \cite{colombier2009complete}. 
In the hole-doped \BaK, the maximum of the superconducting transition temperature \tc \ reaches 38\,K, which is the highest value among those in \BaAs-based superconductors. 

These materials also exhibit the tetragonal-orthorhombic structural transition at or above the antiferromagnetic transition temperature. 
Strong in-plane anisotropy has been observed inside the orthorhombic phase, implying that the structural transition is driven by the lattice coupling to electronic nematicity, which spontaneously breaks the four-fold rotational symmetry of the underlying lattice \cite{chu2010plane}. 
\tc \ becomes maximum near the putative quantum critical point (QCP) of antiferromagnetic and/or electronic nematic order, suggesting that quantum fluctuations associated with the QCP may promote the high-\tc \ superconductivity and lead to the normal state properties deviated from the standard Fermi liquid theory.

High-temperature superconductivity in cuprate superconductors also appears in the vicinity of the antiferromagnetic ordered phase, but there is a crucial difference from iron-based superconductors in the electronic structures of parent compounds.
While undoped cuprates are characterized by a half-filled band and strong Coulomb repulsion drives the system into a Mott insulator, \BaAs \ with 3$d^{6}$ electronic configuration exhibits a metallic ground state.
However, one can approach a half-filled band state in BaFe$_2$As$_2$-based superconductors by hole doping.
In the case of \BaK, quasiparticle mass enhancement toward \KAs \ with 3$d^{5.5}$ is found from electronic specific heat measurements \cite{storey2013electronic}, and an increase of Fe magnetic moment with hole doping is reported by x-ray spectroscopy \cite{Lafuerza2017}.
These results are considered to be connected to the theoretically proposed Mott insulating phase at 3$d^{5}$ configuration \cite{misawa2012ab,medici2014selective}. 
It is also found that the isovalent substitution for K with larger alkali metal ions, Rb or Cs, further enhances electronic correlations, resulting in a very large effective mass, comparable to that of  $f$-electron heavy fermion materials \cite{wang2013calorimetric,mizukami2016,eilers2016straindriven,khim2017a}. 



Hole-doped cuprates exhibit a complicated phase diagram, which involves several competing orders \cite{keimer2015from}. 
In the underdoped regime, charge density wave has been ubiquitously observed \cite{Comin2016}, and there is growing evidence for the electronic nematic order \cite{Hinkov2008,Sato2017,ishida2020divergent}. From the analogy to the liquid crystals, these electronic phases with broken symmetries have been discussed in terms of quantum liquid crystals arising from the doped Mott insulator \cite{kivelson1998electronic}. 
Similarly, in heavily hole-doped iron pnictides, possible charge order has been recently suggested from nuclear quadrupole resonance experiments for RbFe$_2$As$_2$ \cite{Civardi2016}. Furthermore, scanning tunneling spectroscopy \cite{liu2019evidence}, nuclear magnetic resonance \cite{li2016reemergeing,Moroni2019} and elastoresistance measurements \cite{ishida2020novel} reveal a novel form of nematicity in this regime, whose nematic director is aligned along the tetragonal [100] or [010] directions, rotated $45^{\circ}$ from the usual nematicity along [110] or [1$\overline{\rm 1}$0] directions found in other iron-based materials.
Although the alternative interpretation for the elastoresistance data has been suggested \cite{Wiecki2021}, the recent field-angle resolved specific heat measurements have provided the thermodynamic evidence for this diagonal nematicity \cite{Mizukami2021}. 


Detailed studies of the physical properties of materials require sizable and high-quality single crystals. Although single crystals of \BaK \ grown by the flux method \cite{luo2008growth,kihou2010single,kihou2016singlecrystal} have been commonly used to survey the electronic properties of the hole-doped iron pnictides, recently reported novel electronic orders in \RbAs \ and their enhanced electron correlations motivate us the investigation of the physical properties of \BaRb. 
However, single crystal growth of \BaRb \ has not been reported except for the Sn-flux method showing a significant Sn contamination in the Ba site \cite{bukowski2009superconductivity,karpinski2009single}, which can affect their intrinsic properties. 
In this article, we report on the synthesis of \BaRb \ single crystals for a wide doping range ($0.14\le x \le 1.00$) by the FeAs self-flux method, free from the Sn contamination. 
Systematic electrical resistivity measurements reveal that non-Fermi liquid behavior near optimal doping and the mass enhancement toward Rb end ($x=1.00$).



\section{METHODS}
\begin{figure}[t]
	\centering
	\includegraphics[width=1\linewidth]{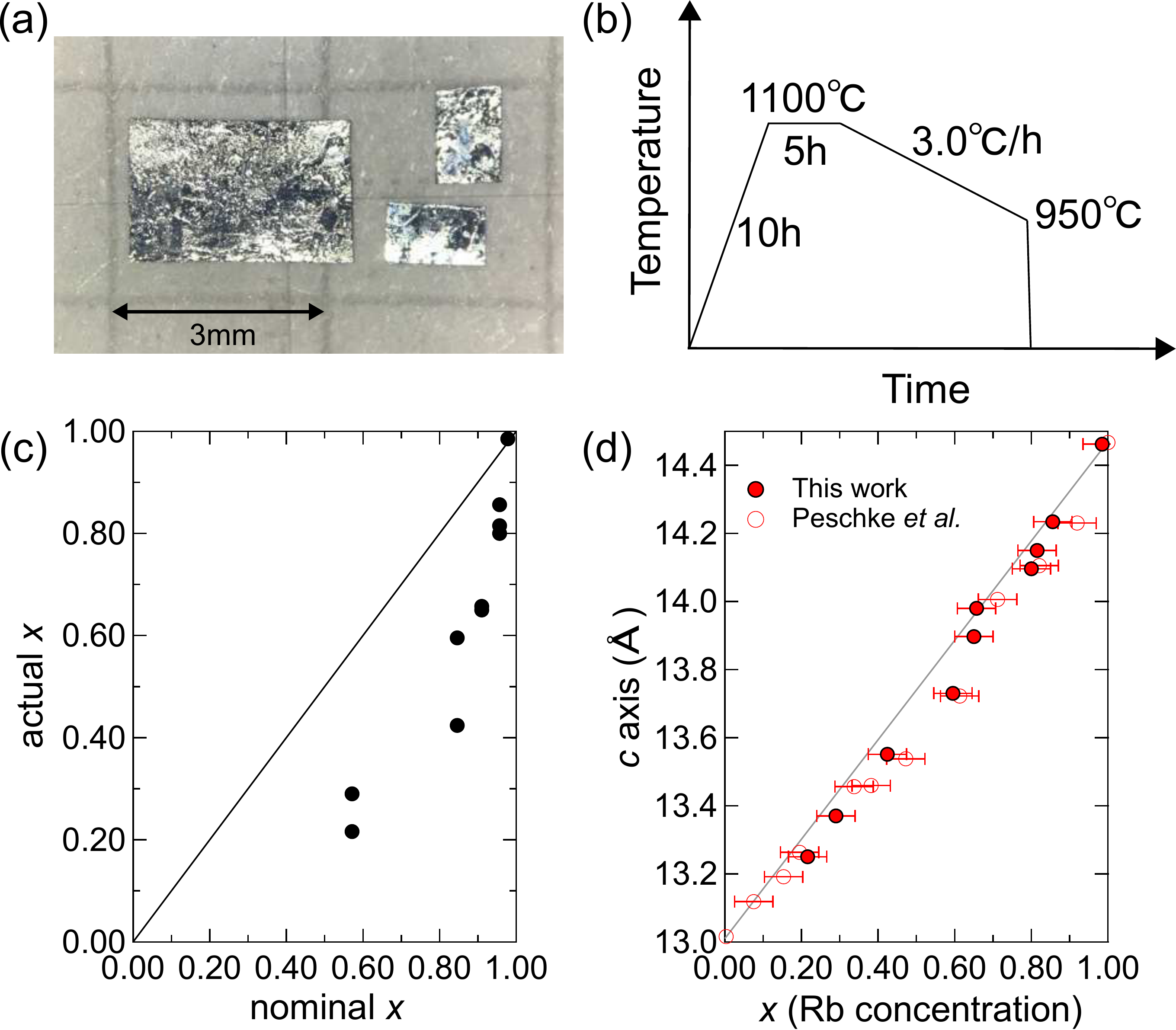}
	\caption{(a) Photographs of \BaRb\ single crystals with $x=0.80$ obtained in this study. (b) Temperature profile of the crystal growth. (c) Nominal $x$ vs actual $x$ determined by EDX. (d) Relationship between the $c$-axis lattice constant determined by X-ray diffraction and actual $x$-value of the single crystals obtained in this study (closed red circles). The gray line represents Vegard's law connecting the data of BaFe$_2$As$_2$ ($c=13.01$\AA) from Ref. \citen{rotter2008sdw} and RbFe$_2$As$_2$ ($c=14.47$\AA) from Ref. \citen{bukowski2010bulk}.
	Open red circles indicate the data of the polycrystalline samples taken from Ref.\citen{peschke2014ba1xrbxfe2as2}.}
	\label{fig1}
\end{figure}

In this study, single crystals of \BaRb \ were synthesized by the FeAs self-flux method.
BaAs, FeAs, and RbAs precursors were prepared in a similar manner as described in Refs. \citen{kihou2010single,kihou2016singlecrystal}.
Starting materials were mixed at appropriate molar ratios, and the mixtures were sealed in evacuated quartz tubes (for preparing BaAs and FeAs) and a stainless steel tube with an alumina crucible (RbAs) and heated up to 700${}^\circ \mathrm{C}$ (BaAs), 900${}^\circ \mathrm{C}$ (FeAs), and 600${}^\circ \mathrm{C}$ (RbAs) for 20 h.
These precursors were weighed in a glove box filled with dried N$_{2}$ gas at the ratios listed in Table\,\ref{table1} and then put into an alumina crucible. The crucible was sealed in a Ta tube or a stainless steel tube using arc welding and loaded into a quartz tube. Then, the quartz tube was welded while evacuating with a rotary pump. 
It was heated up to 1100${}^\circ \mathrm{C}$, kept for 5\,h, and then slowly cooled down to 950${}^\circ \mathrm{C}$ at the rate of $-3^\circ \mathrm{C}$/h, as shown in Fig.\,\ref{fig1}(b). After the crystal growth, the remaining RbAs precursors were rinsed out of the samples and plate-like single crystals were extracted [Fig.\,\ref{fig1}(a)].  

The Rb compositions $x$ of the crystals were determined by energy-dispersive X-ray spectroscopy (EDX) with scanning electron microscopy.  
The crystal structure was evaluated by X-ray diffraction using MoK$\alpha$ radiation (RIGAKU R-AXIS RAPID II).
The dc resistivity measurements were performed by the conventional four-probe method using a a nanovoltmeter (Model 2182A/6221, Keithley) with Delta mode or Physical Property Measurement System (Quantum Design). 
Magnetic susceptibility was measured using a commercial magnetometer (Magnetic Property Measurement System, Quantum Design).

\section{RESULTS AND DISCUSSION}

The ratio of precursors and the resultant Rb concentration $x$ are shown in TABLE\,\ref{table1}, and the relationship between nominal and actual composition is depicted in Fig.\,\ref{fig1}(c). 
Although a positive correlation between nominal $x$ and actual $x$ can be seen, we find pieces of crystals with different Rb contents in the same batch, especially for the low Rb concentration regime.  
Figure\,\ref{fig1}(d) shows the relationship between the $c$-axis lattice constant determined by X-ray diffraction and the $x$-value obtained from the EDX analysis. The error bars represent uncertainties in EDX measurement ($\pm0.05$).
The $c$-axis lattice constant varies almost linearly with Rb concentration, following Vegard's law.

\begin{table}[t]
	\caption{The ratio of precursors (BaAs, RbAs, FeAs) and the resultant Rb concentration $x$. \newline}
	\label{table1}
\begin{center}
	\begin{tabular}{c|c|c|c}
	\hline	
	BaAs  & RbAs & FeAs & $x$ (Rb concentration) \\ \hline
	0.025 & 1.1  & 4    & 0.99                 \\
	0.05  & 1.1  & 4    & 0.80, 0.82            \\
	0.10   & 1.1  & 4    & 0.65, 0.66           \\
	0.20   & 1.1  & 4    & 0.42, 0.60            \\
	0.60   & 0.8  & 4    & 0.14, 0.22, 0.29          \\
	\hline	
	\end{tabular}
\end{center}
\end{table}

\begin{figure}[t]
	\centering
	\includegraphics[width=1\linewidth]{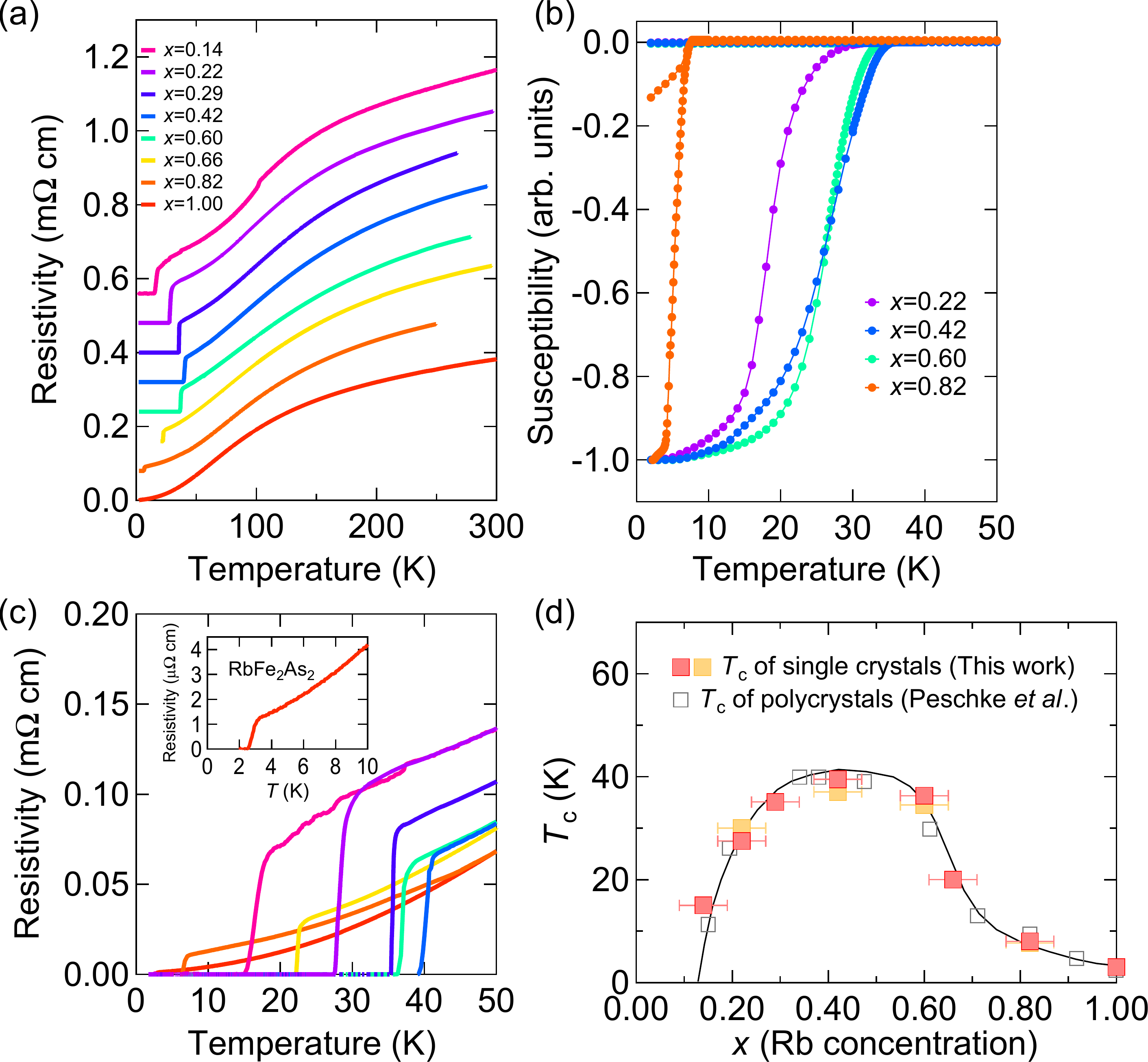}
	\caption{(a) Temperature dependence of in-plane resistivity in single crystals of \BaRb. The data are vertically shifted for clarity. (b) Temperature dependence of magnetic susceptibility measured in a magnetic field of 100\,Oe perpendicular to the $c$-axis in the zero-field-cooling and field-cooling conditions. (c) Resistivity curves near the superconducting transitions. Inset shows the data of \RbAs. (d) Superconducting phase diagram of \BaRb. While the yellow squares represent the onset temperatures of the diamagnetic signals observed in the susceptibility measurements, the red squares indicate \tc \ defined by the zero resistivity criteria. The data for polycrystals in a previous report \cite{peschke2014ba1xrbxfe2as2} are also plotted for comparison (black squares).}
	\label{fig2}
\end{figure}

\begin{figure}[t]
	\centering
	\includegraphics[width=1\linewidth]{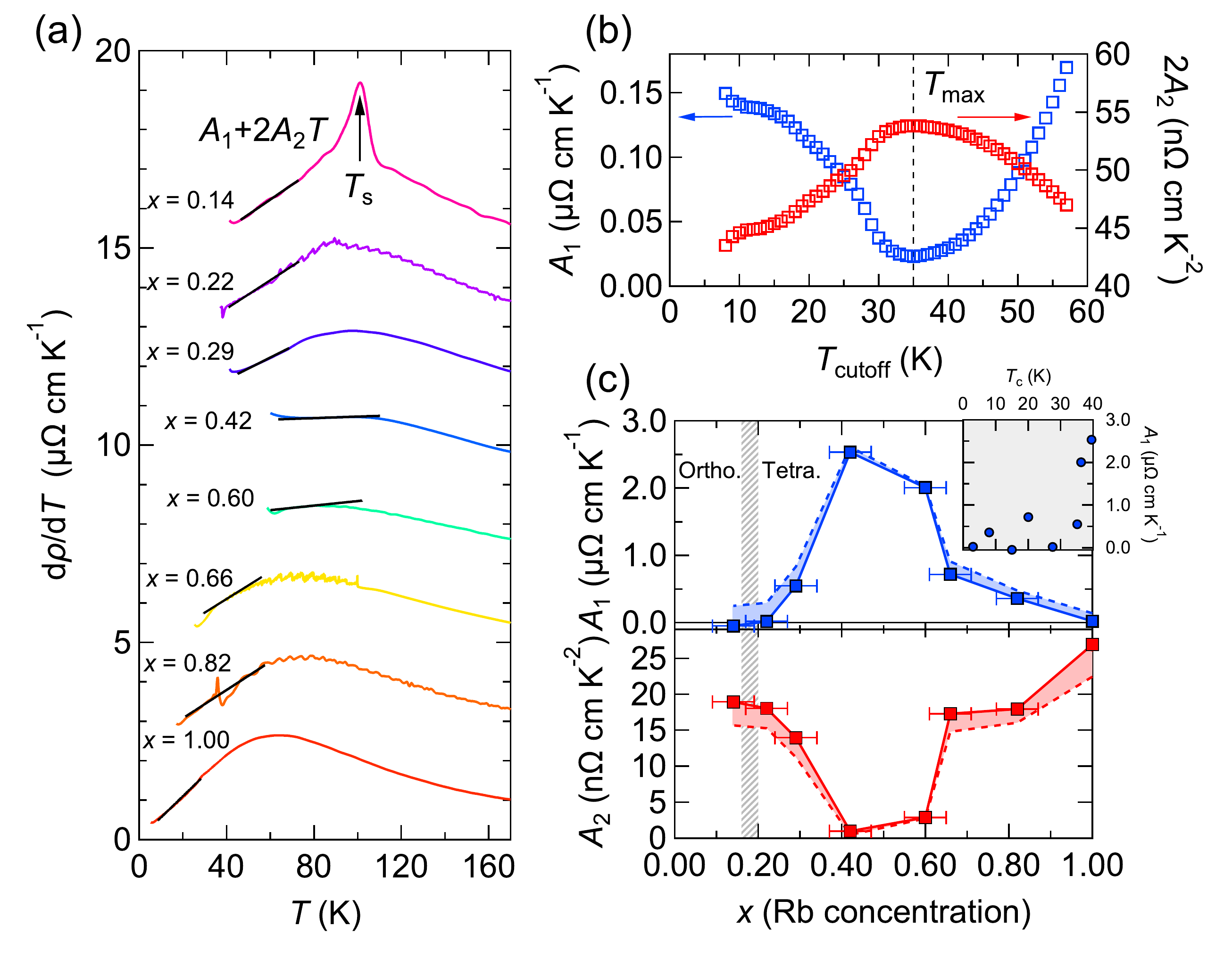}
	\caption{(a) Temperature derivatives of $\rho(T)$ curves shown in Fig.\,\ref{fig2}(a). Black lines represent the linear fits to each data with temperature ranges up to $T_{\rm max}$ ($d\rho/dT=A_1+2A_2T$).
	Only $x=0.14$ shows a clear signature of structural transition, which is denoted by the black arrow. 
	The data are shifted vertically for clarity. 
	(b) Variation of the parameters $A_1$ and $2A_2$ of \RbAs \ when the upper bound of fitting range $T_{\rm cutoff}$ varies.
	(c) $x$-dependence of the fitting parameters $A_1$ and $A_2$. Closed squares represent the values obtained when $T_{\rm cutoff}=T_{\rm max}$, and the dashed lines indicate their possible maximum/minimum values estimated in their $T_{\rm cutoff}$ dependence. 
	The shades show the range of possible values.
	Inset shows the $A_1$ values plotted against their \tc.}
	\label{fig3}
\end{figure}

The temperature dependence of in-plane resistivity $\rho (T)$ is shown in Fig.\,\ref{fig2}(a). 
A clear superconducting transition is detected in all the samples [Fig.\,\ref{fig2}(c)]. 
In \RbAs, the residual resistivity ratio $RRR=\rho(300\rm{K})/\rho_0$, where $\rho_0$ is the residual resistivity extrapolated to $T\to 0$ by the power-law fitting (see below), is about 250, indicating the high quality of our crystals. 
The onset temperatures of the diamagnetic signals measured by dc magnetic susceptibility [Fig.\,\ref{fig2}(b)] are in line with \tc\, determined by the zero resistivity [Fig.\,\ref{fig2}(d)].
Moreover, the recent specific heat study \cite{Mizukami2021} confirmed the bulk superconductivity in the grown crystals.
The superconducting phase diagram is shown in Fig.\,\ref{fig2}(d). 
The doping dependence of the superconducting transition temperature \tc$(x)$ in single crystals obtained in the present study coincides well with that of polycrystals reported previously \cite{peschke2014ba1xrbxfe2as2}.
The corresponding temperature derivatives of the resistivity curves depicted in Fig.\,\ref{fig3}(a) exhibit a discernable anomaly associated with a structural/magnetic transition only in $x=0.14$, implying that the boundary of the orthorhombic phase locates between $x=0.14$ and $x=0.22$.

To study the nature of the quasiparticle scattering in \BaRb, we analyze the temperature dependence of the electrical resistivity. 
First, we show the results of polynomial fitting for the resistivity curves ($\rho(T) = \rho_0 + A_1 T + A_2 T^2$). 
Since the overall $\rho(T)$ curves show saturating behaviors at high temperatures (see Fig.\,\ref{fig2}(a)), the sufficiently low-temperature region should be used for the fitting analysis. 
However, due to the strong doping dependence of \tc, the fitting range has to be set depending on the doping level.
To obtain $A_1$ and $A_2$ values through the reasonable analysis, here we focus on the $d\rho/dT$ data shown in Fig.\,\ref{fig3}(a).
We perform a linear fitting to the temperature derivative of resistivity as $d\rho/dT=A_1+2A_2T$ from above \tc \ to the upper bound $T_{\rm cutoff}$, varied up to the temperature at which the $d\rho/dT$ curves show a hump feature. 
Figure \,\ref{fig3}(b) shows the parameters $A_1$ and $2A_2$ versus $T_{\rm cutoff}$ in \RbAs.
As $T_{\rm cutoff}$ becomes higher, $2A_2$ first increases and then turns to decrease at $T_{\rm max}$, above which $d\rho/dT$ starts to deviate from a linear form. 
Qualitatively similar $T_{\rm cutoff}$ dependence of $A_1$ and $2A_2$ are seen in all the compositions. 
In the upper (lower) panel of Fig.\,\ref{fig3}(c), the $x$ dependence of $A_1$ ($A_2$) values at $T_{\rm cutoff}=T_{\rm max}$ are plotted as closed squares together with their possible maximum (minimum) values found in the $T_{\rm cutoff}$ dependence of $A_1$ and $2A_2$ as a dashed line. 
Although there are uncertainties in their precise values, their doping evolutions $A_1(x)$ and $A_2(x)$ exhibit a clear trend, which does not depend on $T_{\rm cutoff}$.
Non-Fermi liquid $T$-linear contribution is enhanced at the optimal compositions $x=0.42$ and $x=0.60$, and as moving away from this regime, $A_1$ decreases and the Fermi-liquid $T^2$ component becomes larger.
Indeed, this trend itself is already visible in the $d\rho/dT$ data without the above fitting procedures: $d\rho/dT$ traces of $x=0.42$ and $x=0.60$ are flatter than those of other compositions. 


Similar second-order polynomial fits as used here have successfully described the temperature dependence of resistivity in organic superconductors and electron-doped iron pnictides \BaCo \ \cite{doironleyraud2009correlation}. 
In these materials, the $T$-linear behavior of  resistivity becomes most pronounced near the antiferromagnetic QCP, and away from the QCP, it recovers the standard $T^2$ dependence expected in conventional metals. 
In the crossover region, the temperature dependence of resistivity can be expressed as a sum of these components, and the coefficient of the $T$-linear component has been found to scale with \tc. 
In our results, however, the $T$-linear coefficient $A_1$ shows a broad maximum around the optimal doping, while the endpoint of the antiferromagnetic phase locates in the underdoped regime $x\sim 0.20$, where the antiferromagnetic transition temperature goes to zero \cite{peschke2014ba1xrbxfe2as2}. 
Furthermore, as shown in the inset of Fig.\,\ref{fig3}(c), the $T$-linear term $A_1(x)$ does not seem to scale perfectly linearly with \tc. 
These results are not compatible with the description that non-Fermi liquid behavior near  antiferromagnetic QCP correlates with \tc \ \cite{taillefer2010scattering}, but this situation is similar to the case of \BaK \ \cite{liu2014comprehensive}, in which the enhanced $T$-linear term can be found at $x\simeq 0.40$, away from the endpoint of the antiferromagnetic phase.

\begin{figure}[t]
	\centering
	\includegraphics[width=1\linewidth]{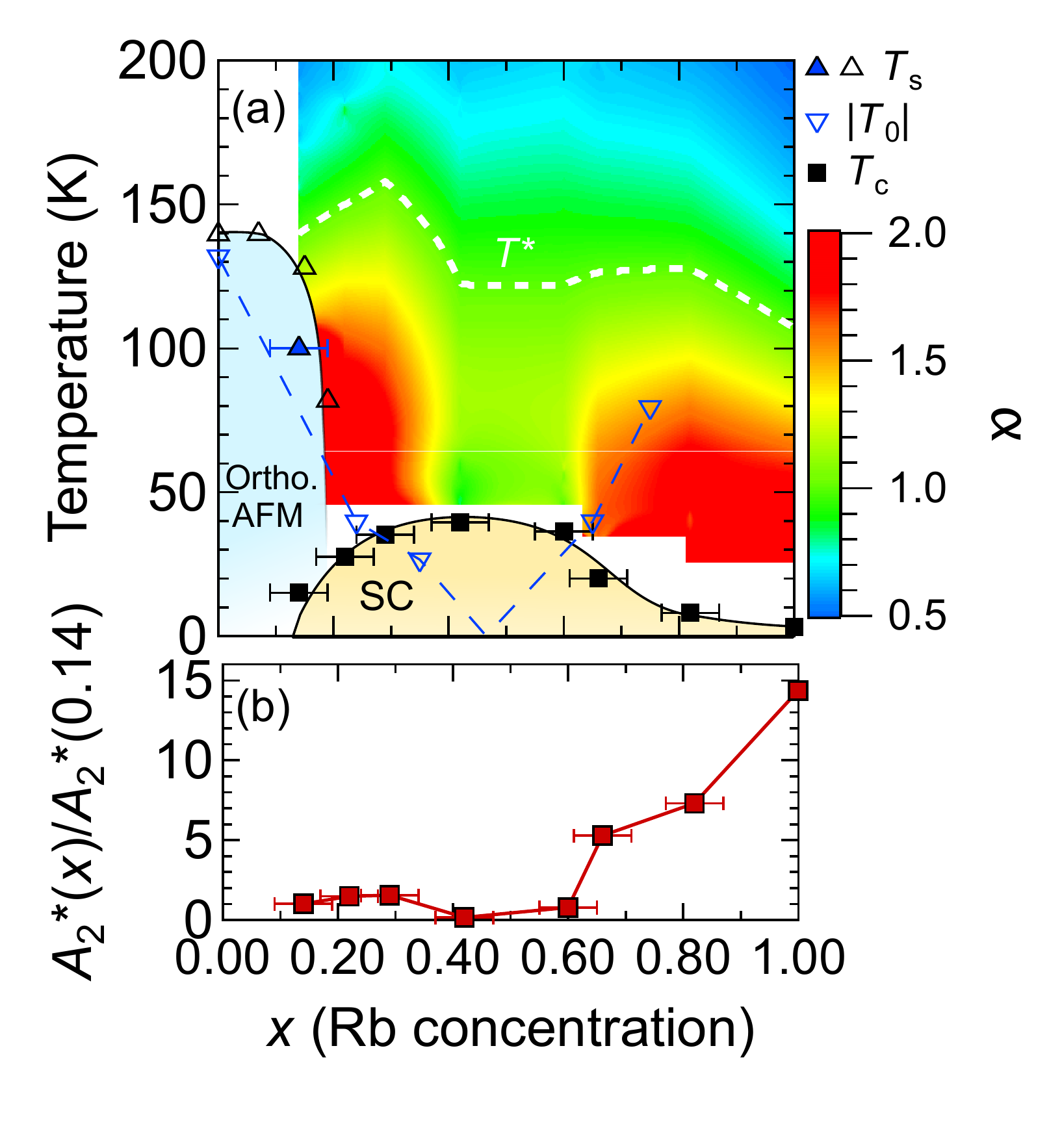}
	\caption{(a) Phase diagram of \BaRb \ with color contour of exponent $\alpha$ obtained by the power-law fitting analysis ($\rho(T) = \rho_0 + AT^{\alpha}$). Open triangles represent the structural transition temperature $T_{\rm s}$ from ref.\citen{peschke2014ba1xrbxfe2as2}, and $T_{\rm s}$ \ of $x=0.14$ detected in our grown single crystal is shown as a closed triangle. 
	Closed squares indicate \tc \ of single crystals determined by the resistivity data. 
	Open inverse triangles show the absolute value of the Curie-Weiss temperature $|T_0|$ in the nematic susceptibility \cite{ishida2020novel}. 
	White dashed line represents $T^{*}$, above which $\rho(T)$ shows convex up curvature ($\alpha<1$).
	(b) Doping dependence of modified $T^2$ coefficient $A_2^*$ normalized by its value for $x=0.14$.}
	\label{fig4}
\end{figure}

Alternatively, the $\rho(T)$ data can also be analyzed by the power-law fitting, $\rho(T) = \rho_0 + AT^{\alpha}$.  
In Fig.\,\ref{fig4}(a), we map the temperature evolution of the exponent $\alpha$ obtained by such an analysis. 
We use $\rho_0$ extracted by the power fitting between above \tc \ to $T_{\rm max}$, and $\alpha$ was calculated by a linear fit to $\ln{(\rho-\rho_0)} = \alpha\ln{T} + \ln{A}$. 
The fitting width was set to 5\,K and the fitting range was slid by 1\,K. 
The results of this power-law analysis also show that the sublinear temperature dependence with $\alpha \sim 1.1$ emerges around $x=0.42-0.60$.
With decreasing or increasing composition $x$ from this region, the low-temperature resistivity shows a crossover behavior from the sublinear to quadratic $T$-dependence. 
This trend is consistent with the results of the polynomial fitting.

The color plot of the exponent $\alpha$ in Fig.\,\ref{fig4}(a) constructs a fan shape centered around the optimal composition, where the sublinear temperature dependent resistivity appears. 
This reminds us of the phase diagram with an antiferromagnetic quantum critical point, as found in the isovalently substituted iron pnictide BaFe$_2$(As$_{1-x}$P$_x$)$_2$ \cite{kasahara2010} and heavy-fermion compound YbRh$_2$Si$_2$ \cite{custers2003break}. 
However, as we have already mentioned, in our case of \BaRb, the antiferromagnetic phase fades away around $x\sim0.20$, far from the center of the fan, and thus the observed fan shape is unlikely related to the antiferromagnetic quantum criticality. 

At present, the origin of the non-Fermi liquid sub-$T$-linear behavior observed far away from the antiferromagnetic endpoint is not clear. 
However, we point out that this may be related to electronic nematic instability. 
There is growing experimental evidence that quantum fluctuations of nematic order can give rise to non-Fermi liquid properties \cite{Licciardello2019,Licciardello2019a,Huang2020}, especially near the nematic quantum critical point found in FeSe$_{1-x}$S$_x$ \cite{Hosoi2016,Ishida2022}.  
Recent elastoresistivity measurements in the \BaRb\ system have revealed that the nematic susceptibility $\chi_{\rm nem}(T)$ in the underdoped side shows the Curie-Weiss temperature dependence $\sim (T-T_0)^{-1}$ with a positive Curie-Weiss temperature $T_0 (> 0)$ even in the doping range outside the orthorhombic phase \cite{ishida2020novel}. 
The Landau free-energy analysis indicates that the Curie-Weiss temperature $T_0$ corresponds to the bare nematic transition temperature with no coupling between the electronic system and the lattice \cite{chu2012divergent}. 
In real materials, the presence of nemato-elastic coupling leads to the increase of actual nematic (structural) transition temperature $T_{\rm s}$ from $T_0$, i.e. $T_{\rm s}>T_0$ \cite{Paul2017}. 
Therefore the positive $T_0$ observed in the tetragonal phase with no $T_{\rm s}$ implies that there is some mechanism that suppresses the structural transition although the electronic system has a tendency toward nematic instability. 
Indeed, recently, it is theoretically pointed out that nematic susceptibility close to the nematic QCP can show a deviation from the a Curie-Weiss law at low temperatures when it is well separated from magnetic QCP \cite{Tazai2022}.
On the other hand, a similar situation can also be found in the hole-doped \BaK\ system, in which positive $T_0$ values have been reported near the optimal composition with no structural transition \cite{Kuo2016,Terashima2020}. 
In \BaK, it has been reported that near the endpoint of the magnetic phase, there exists a phase transition from the $C_2$ stripe-type antiferromagnetic phase to a $C_4$ magnetic phase \cite{Bohmer2015}, implying the existence of competing magnetic instabilities. 
Such a $C_2$--$C_4$ competition would prevent the nematic order even in the region of the phase diagram where the bare nematic transition temperature $T_0$ is finite. 
In \BaRb, the reported systematic trends of $T_0$ in Ref.\,\citen{ishida2020novel} reveal a sign change between $x\sim 0.40$ and $0.65$, as depicted in Fig.\,\ref{fig4}(a).
Taking into account this situation, it is tempting to relate the observed strange metal component to the hidden quantum critical fluctuations of nematic instability fingerprinted by $T_0\to 0$, but their high superconducting transition temperature prevents us from seeing the precise form of their elastoresistivity down to zero temperature.
Measurements under high magnetic fields would be helpful to verify this point. 



Next, we discuss the doping dependence of the effective mass of quasiparticles, which is related to the $AT^2$ dependence of resistivity, a hallmark of Landau's Fermi liquid theory. 
The $AT^2$ resistivity comes from the constraints on the phase space where two electrons near the Fermi surface take part in the electron-electron scattering event. The coefficient $A$ is proportional to $E_{\rm F}^{-2}$, where $E_{\rm F}$ is Fermi energy, and thus this coefficient gives a measure of the electron correlations. It has also been widely discussed in terms of the Kadowaki-Woods relation \cite{kadowaki1986universal} that $A$ is proportional to the square of the electronic specific heat coefficient $\gamma^2$, which is a measure of quasiparticle effective mass.


In a multiband quasi-2D metal with cylindrical Fermi surfaces, the $A$ coefficient is given as \cite{hussey2005,Licciardello2019}:
\begin{equation}
A=\frac{8\pi ack^2_{\rm B}}{e^2\hbar^3}\frac{1}{\sum_i k_{{\rm F}_i}^3/m_i^{*2}},
\label{eqA}
\end{equation}
where $a$ ($c$) is the $a$-axis ($c$-axis) lattice parameter, $k_{\rm{F}_i}$ is the Fermi wave vector, and $m_i^{*}$ is the effective mass for each Fermi pocket $i$.
The Fermi wave vector for a cylindrical Fermi surface is expressed as $k_{{\rm F}_i} = \sqrt{2\pi cn_i}$ ($n_i$ is the carrier density), and thus $A$ depends on $m_i^*$ and $n_i$ as $A \propto \left(\sum_i n_i^{\frac{3}{2}}/m_i^{*2}\right)^{-1}$. 
From the carrier density and the quasiparticle effective mass of \RbAs \ summarized in Ref. \citen{hardy2016strong}, we can estimate $A \sim 36$\,n\textohm\,cm\,$\rm{K}^{-2}$, which is in a reasonable agreement with the measured value $A_2 \sim27$\,n\textohm\,cm\,$\rm{K}^{-2}$ [Fig.\,\ref{fig3}(b)]. 
This implies that the $T^2$ resistivity in \RbAs \ can be attributed to the electron-electron scattering. 
With the value of electronic specific heat coefficient $\gamma \sim120$ mJ $\rm{mol}^{-1}$ $\rm{K}^{-2}$ of \RbAs, recently measured in the single crystals obtained in the present work \cite{Mizukami2021}, this leads to the large Kadowaki-Woods ratio $A_2/\gamma^{2} \sim 1.88 \times 10^{-6}$\,\textmu\textohm\,cm\,$\rm{K}^{2} \rm{mol}^{2} \rm{mJ}^{-2}$, demonstrating its strong correlation.
To discuss the general doping trend of effective mass, in the following we use the coefficients $A_2$ in the second-order polynomial fits, which are close to the results of the simple $AT^2$ fits except for $x\sim 0.42$ and $0.60$, where the $T$-linear behavior is overwhelming. 

As we see above, coefficient $A$ depends not only on the effective mass but also on the carrier density, which should be taken into account to discuss the evolution of $A$ with hole doping.
As indicated in Eq.\,\ref{eqA}, if $m^{*}$ were independent of Rb concentration, $A$ would decrease as carrier density increases.
However, as shown in Fig.\ref{fig3}(b), $A_2$ is found to increase slightly with increasing $x$ in the high doping region. 
To examine the doping evolution of $m^{*}$ from $A_2$, we define a new parameter $A_2^*(x)$ as $A_2(x)\times n_{\beta}^{\frac{3}{2}}(x)$.
Here, we discuss the doping dependence of carrier density as represented by the carrier density $n_{\beta}$ in the outer hole sheet ($\beta$ band), which is reported to show a significant mass enhancement by hole doping and become more than twice as large as other sheets in \RbAs \ \cite{hardy2016strong}.
To estimate $n_{\beta}(x)$, we simply assume that the $x$ dependence of carrier density in \BaRb \ is the same as \BaK, and $n_{\beta}(x)$ linearly changes with $x$. 
By using the data of $n_{\beta}(0.4)$ and $n_{\beta}(1)$ for \BaK \ summarized in Ref. \citen{hardy2016strong}, the doping evolution of $A_2^*(x)$ normalized by the $x=0.14$ value is obtained as shown in Fig.\,\ref{fig4}(b). Although the above assumptions have some quantitative uncertainties, it is unmistakable that $A^*(x)$ grows rapidly as the Rb concentration approaches the high doping end, which suggests that the effective mass of quasiparticles is largely enhanced toward $x=1.00$.


The increasing trend of $A_2^*(x)$ with $x\to1.00$ is compatible with the scenario that the system approaches the $3d^5$ half-filled Mott insulating state. 
In contrast to high-\tc \ cuprates, the driving force of electron correlations in iron pnictides is not on-site Coulomb repulsion but the Hund's coupling effect, which makes the different $d$ orbitals decoupled \cite{medici2014selective}.
This leads to the coexistence of the localized and iterant $d$ orbitals, and a crossover from the incoherent state to coherent heavy Fermi liquid with lowering temperature \cite{hardy2013evidence}.
This incoherent-coherent crossover behavior can be seen in the temperature dependence of resistivity in \BaRb\ [Fig.\,\ref{fig2}(a)], where the curvature of $\rho(T)$ changes from convex upward ($\alpha<1.0$) at high temperatures to convex downward ($\alpha>1.0$) at low temperatures. 
We can define $T^*$ at which this curvature change occurs ($\alpha=1.0$), and plot $T^*(x)$ in Fig.\,\ref{fig4}. 
We find that $T^{*}$ decreases with Rb concentration $x$, indicating the suppression of crossover temperature, which is consistent with the enhanced correlations with hole doping, as discussed in the $f$-electron Kondo systems \cite{yang2012emergent}.

\section{CONCLUSIONS}

To sum up, we have synthesized a series of single crystals of \BaRb \ with $0.14 \leq x \leq 1.00$ by the FeAs self-flux method. 
The lattice constant follows Vegard's law, and magneto-structural transition disappears between $x=0.14$ and $0.22$, although the positive bare nematic transition temperature $T_0$ from the elastoresistivity measurements is reported even for $x\sim 0.40$, implying that there is some mechanism to prevent $C_2$ nematic ordering. 
By analyzing in-plane electrical resistivity $\rho(T)$ via polynomial and power-law fits, the non-Fermi liquid $T$-linear contribution is found to become pronounced around the optimal composition, away from the antiferromagnetic endpoint.
The analysis of the Fermi-liquid coefficient of the resistivity curve indicates that the effective mass $m^*$ rapidly grows near the high doping end, consistent with the putative Mott-insulating phase near the $3d^5$ half-filled state. 

\section*{Acknowledgements}

We thank fruitful discussion with M. Nakajima and M. Tanatar. 
This work was supported by Grants-in-Aid for Scientific Research (KAKENHI) (No.\ JP18H05227, No.\ JP19H00649, No.\ JP20K21139, No.\ JP20H02600), Grants-in-Aid for Scientific Research on innovative areas ``Quantum Liquid Crystals” (No.\ JP19H05823, No.\ JP19H05824) from Japan Society for the Promotion of Science (JSPS).

\bibliography{ref.bib}

\end{document}